\documentclass[namedreferences,hyperref,optionalrh]{spr-sola}
\usepackage{geometry}
\usepackage{graphicx}        
\usepackage{color} 
\usepackage{mathtools}

\usepackage{amsfonts,amsmath,amssymb}

\newcommand{\araa}{{\it Ann. Rev. Astron. Astrophys.}} 
 
\newcommand{\aap}{{\it Astron. Astrophys.}}

\newcommand{\apj}{{\it Astrophys. J.}}
\newcommand{\apjl}{{\it Astrophys. J. Lett.}}

\newcommand{\mnras}{{\it Mon. Not. R. Astron. Soc.}}

\newcommand{\solphys}{{\it Sol. Phys.}}

\chardef\us=`\_

\def\<{\,\langle\langle}
\def\>{\,\rangle\rangle}

\def\avg#1{\langle#1\rangle}

\defcitealias{2019ApJ...875..158W}{WP19}

\renewcommand{\v}[1]{\ensuremath{\mathbf{#1}}} 
\newcommand{\gv}[1]{\ensuremath{\mbox{\boldmath$ #1 $}}} 
\newcommand{\pd}[2]{\frac{\partial #1}{\partial #2}}
\newcommand{\pdtext}[2]{\partial #1/\partial #2}
\newcommand{\f}{\frac}  

\usepackage{letltxmacro,amsmath}
\LetLtxMacro{\originaleqref}{\eqref}
\renewcommand{\eqref}{Eq.~\originaleqref}
\newcommand{\ineqref}{condition~\originaleqref}
\begin{document}

\begin{frontmatter}
\title{Catastrophic cooling instability in optically thin plasmas}

\author[addressref={aff1,aff2}]{\inits{T.}\fnm{Timothy}~\snm{Waters}\orcid{0000-0002-5205-9472}}
\author[addressref={aff1,aff3},corref,email={astricklan@lanl.gov}]{\inits{A.}\fnm{Amanda}~\snm{Stricklan}\orcid{0000-0002-5205-9472}}
\address[id=aff1]{X-Computational Physics Division, Los Alamos National Laboratory}
\address[id=aff2]{Center for Theoretical Astrophysics, Los Alamos National Laboratory, Los Alamos, NM, USA}
\address[id=aff3]{Department of Astronomy, New Mexico State University P.O. Box 30001, MSC 4500 Las Cruces, NM 88003-8001, USA}

\runningauthor{Waters \& Stricklan}
\runningtitle{Catastrophic Cooling Instability}

\begin{abstract}
The solar corona is the prototypical example of a low density environment heated to high temperatures by external sources.  
The plasma cools radiatively, and because it is optically thin to this radiation, it becomes possible to model the density, velocity, and temperature structure of the system by modifying the MHD equations to include an energy source term that approximates the local heating and cooling rates.  
The solutions can be highly inhomogeneous and even multiphase because the well known linear instability associated with this source term, thermal instability, leads to a catastrophic heating and cooling of the plasma in the nonlinear regime.  
Here we show that there is a separate, much simpler linear instability accompanying this source term that can rival thermal instability in dynamical importance. 
The stability criterion is the isochoric one identified by \citet{1953ApJ...117..431P}, and we demonstrate that cooling functions derived from collisional ionization equilibrium are highly prone to violating this criterion.   
If catastrophic cooling instability can act locally in global simulations, then it is an alternative mechanism for forming condensations, and due to its nonequilibrium character, it may be relevant to explaining a host of phenomena associated with the production of cooler gas in hot, low density plasmas.
\end{abstract}
\end{frontmatter}

\section{Introduction}

Catastrophic cooling refers to a runaway radiative cooling of gas that proceeds so long as no balance with heating processes is reached.   
There will be an accompanying loss of pressure support as the thermal state of the gas departs further away from equilibrium, hence mechanical equilibrium will also be lost.  
A widely studied example of this occurs in the (multi-)million degree plasma within coronal loops \citep[for recent reviews, see][]{2020PPCF...62a4016A,2020RAA....20..166C,2022FrASS...920116A}.
The descriptor `catastrophic cooling' has been used to characterize the formation mechanism of both prominences and coronal rain \citep[e.g.][]{1999ApJ...512..985A,2001SoPh..198..325S,2001ARA&A..39..175A,2004A&A...424..289M,2005A&A...436.1067M,2010ApJ...716..154A,2012A&A...537A.152P, 2024arXiv240805736Q, 2024ApJ...967...82R}, as well as the bulk cooling observed in numerical simulations of pulse-heated loops \citep[e.g.][]{2005ApJ...624.1080M, 2012A&A...543A..90R,2013ApJ...772...40C,2024ApJ...965...53K}.
Most multi-dimensional simulation studies of coronal loops attribute coronal rain to be the outcome of local thermal instability (TI) \citep{2013ApJ...771L..29F,2015ApJ...807..142F, 2015AdSpR..56.2738M, 2022ApJ...926L..29A,2023MNRAS.526.1646D,2024ApJ...971...90D}, a linear instability of non-adiabatic gas dynamics that was first elucidated by \citet{1965ApJ...142..531F}.  

The nonlinear regime of TI is itself a catastrophic cooling process, although one in which the density can rise by nearly the same factor that the temperature can drop.   
The way in which thermal equilibrium is regained as TI saturates is commonly covered in textbooks and amounts to drawing an isobar on a pressure-density phase diagram connecting different points where heating can balance cooling \citep[e.g.][]{1992phas.book.....S}.  
Only recently was the mechanism for restoring mechanical equilibrium shown to be damped oscillations mediated by sound waves \citep[][hereafter referred to as \citetalias{2019ApJ...875..158W}]{2019ApJ...875..158W}; \citet{2023FrASS..1098135W} confirmed this analytically and illustrated the limitations of the simple isobar analysis.
Coincident observations of long-period intensity pulsations being in phase with periodic episodes of coronal rain might be regarded as evidence of coronal plasma `struggling' to regain mechanical equilibrium upon cooling to a new equilibrium temperature\citep{2018ApJ...853..176A}. 

The observed properties of these pulsations appear to instead lend support to the thermal nonequilibrium (TNE) model \citep{2015ApJ...807..158F, 2017ApJ...835..272F, 2018ApJ...855...52F}, which is an alternative scenario --- specific to coronal loops --- for forming condensations.  
The TNE model distinguishes itself from a broader class of evaporation-condensation models \citep[e.g.][]{1979SoPh...64..267P, 1986SoPh..104..303P, 1990ApJ...359..228M} through a heating deposition that is concentrated at the footpoints of a loop \citep{1991ApJ...378..372A, 1999ApJ...512..985A, 2000ApJ...536..494A, 2005ApJ...635.1319K,2021ApJ...916..115S}.   
Despite TNE being a fully nonlinear and cyclic condensation process, it is still up for debate whether gas condenses at some point in the cycle due to the growth of unstable linear modes \citep[e.g.][]{2019SoPh..294..173K}. 
In support of the argument to disfavor TI is the fact that the local approximation will often not hold in coronal loops and, more importantly, that both the effects of line-tying and thermal conduction should stabilize TI along the magnetic field for all but the longest loops \citep[e.g.][]{1979SoPh...64..267P,1984ApJ...276..755A,1984ApJ...284..422A}.
In simulations adopting the TNE heating prescription, however, coronal rain is still interpreted as being a result of TI \citep[e.g.][]{2015AdSpR..56.2738M,2022ApJ...926L..29A}. 
Adding to the controversy, \citet{2015ApJ...807..142F} describe TNE as simply the nonlinear counterpart of TI, while \citet{2022FrASS...920116A} refer to the condensation process leading to either prominences or coronal rain as `TNE-TI cycles'.
\par

Here we suggest that this contentious state of affairs has arisen because it has not been recognized that, separate from TI, there is an elementary linear theory solution to the equations of non-adiabatic MHD describing catastrophic heating and cooling. 
In \S{2}, we present the linear theory of this instability and derive a bound for when it can overtake TI when both instabilities operate simultaneously.  
In \S{3}, we demonstrate that commonly used radiative loss functions in solar physics, those representing collisional ionization equilibrium, are highly prone to violating the isochoric stability criterion that governs the eigenmode of this instability.  
In \S{4}, we describe how these findings are relevant to resolving the above controversy.  
In the Appendices, we address the subtlety of this instability, highlighting how several authors have performed similar analyses without concluding that there is a separate eigenmode that can act independently of TI.  
\par

\section{\label{sec:2}Linear Theory}
The net cooling function, $\mathcal{L} = \mathcal{L}(\rho,T)$, representing losses minus gains in units of $\rm{erg\,g^{-1}\,s^{-1}}$, enters the internal energy equation for non-adiabatic, ideal MHD as the volumetric source term $-\rho\mathcal{L}$: 
\begin{equation}
   \rho\f{D\mathcal{E}}{Dt} = -p\mathbf{\nabla} \cdot \gv{v} - \rho \mathcal{L} - \mathbf{\nabla} \cdot \gv{q}.
   \label{eq:energy}
\end{equation}
Here, the variables $\rho$, $\gv{v}$, and $p$ are the gas density, velocity, and pressure, respectively, $\mathcal{E} = c_V T$ is the gas internal energy, $\gv{q}$ is the heat flux due to thermal conduction, and $D/Dt$ is the comoving frame derivative.
In the rest frame of a homogeneous plasma (i.e. one with $\mathbf{\nabla} \cdot \gv{v} = 0$ and $\mathbf{\nabla} \cdot \gv{q} = 0$) with a constant specific heat capacity $c_V$, \eqref{eq:energy} becomes simply
\begin{equation}
   c_V\pd{T}{t} = -\mathcal{L}.
   \label{eq:Teqn}
\end{equation}
\par

The plasma reaches its thermal equilibrium temperature $T_{eq}$ when the heating and cooling rates balance ($\mathcal{L} = 0$).  
In the linear theory of TI, \eqref{eq:Teqn} governs the background equilibrium state but the linearized version of \eqref{eq:energy} governs the evolution of perturbations.  
Because that equation involves $\delta\rho$, $\delta \gv{v}$, and $\delta p$, obtaining a dispersion relation requires linearizing the mass and momentum conservation equations also, which in turn requires linearizing the induction equation and the equation of state to obtain a closed set of equations.  
What has evidently been overlooked is that there is in fact an almost trivial solution requiring only the linearized version of \eqref{eq:Teqn}: uniform temperature fluctuations $\delta T$, defined (for an ideal gas with $p = \rho kT/\bar{m}$) as perturbations satisfying $(\delta \rho, \delta \gv{v}, \delta p, \delta \gv{B}) = (0,\gv{0},\rho k \delta T/\bar{m},\gv{0})$, with $\delta T = \delta T(t)$ being spatially constant.
This implies that no flows can develop as $\delta T$ is applied because both the gas density and pressure remain spatially constant.
\par 

\subsection{Catastrophic heating/cooling instability}
Letting $T = T_{eq} + \delta T$ and noting that $\mathcal{L}(T_{eq}) = 0$ by definition,
we have for $|\delta T| \ll T_{eq}$, 
\begin{equation}
\mathcal{L}(T_{eq} + \delta T) = \left(\pd{\mathcal{L}}{T}\right)_\rho \delta T.
\label{eq:Taylor}
\end{equation}
\eqref{eq:Teqn} becomes 
\begin{equation}
   \pd{\ln\delta T}{t} = -\f{1}{c_V}\left(\pd{\mathcal{L}}{T}\right)_\rho,
   \label{eq:deltaTeqn}
\end{equation}
which is an evolution equation for how $\delta T$ changes with time.  The linear theory solution is found by taking $\left(\pdtext{\mathcal{L}}{T}\right)_\rho$ to be time-independent, giving 
\begin{equation}
   \delta T(t) = (\delta T)_0\exp\left[-\f{1}{c_V}\left(\pd{\mathcal{L}}{T}\right)_\rho t \right],
\label{eq:deltaToft}
\end{equation}
where $(\delta T)_0$ is the initial temperature fluctuation above or below $T_{eq}$.
Thus, the background equilibrium state is unstable to either catastrophic cooling, $(\delta T)_0 < 0$, or catastrophic heating, $(\delta T)_0 > 0$, whenever 
\begin{equation}
 \left(\pd{\mathcal{L}}{T}\right)_\rho < 0.
\label{eq:isochoric}
\end{equation}
This instability criterion is just the well known isochoric criterion for TI; see \citet{1965ApJ...142..531F}.  
\par

Because the isochoric derivative is central to this result, we should emphasize that in \eqref{eq:Taylor}, the $\rho$-subscript appears because we are assuming a temperature displacement along a vertical path on a $T$-vs-$\rho$ phase diagram. 
Alternatively, we can simply apply the Eulerian perturbation operator $\delta$ to \eqref{eq:Teqn}, to give
\begin{equation}
   \pd{\delta T}{t} = -\f{\mathcal{\delta L}}{c_V}.
   \label{eq:delta-Teqn}
\end{equation}
With $\mathcal{L} = \mathcal{L}(\rho,T)$, we have that 
\begin{equation}
\mathcal{\delta L} = \left(\pd{\mathcal{L}}{T}\right)_\rho \delta T + \left(\pd{\mathcal{L}}{\rho}\right)_T \delta \rho.
\label{eq:deltaL}
\end{equation}
Substituting \eqref{eq:deltaL} into \eqref{eq:delta-Teqn} and considering solutions with $\delta \rho = 0$ also gives \eqref{eq:deltaTeqn}.
\par

It is essential for our later discussion to notice that this instability is present when including a thermal conductivity, $\kappa(T)$; for constant $\kappa$, for example, \eqref{eq:deltaTeqn} reads
\begin{equation}
   \pd{\delta T}{t} = -\f{1}{c_V}\left(\pd{\mathcal{L}}{T}\right)_\rho \delta T + \f{\kappa}{\rho c_V} \nabla^2 \delta T.
   \label{eq:deltaTeqnTC}
\end{equation}
Obviously, \eqref{eq:deltaToft} is a solution to this equation.
We have chosen not to present the derivation in this manner because it was \citet{1953ApJ...117..431P}'s inclusion of thermal conduction that invited criticism.  
We summarize the TI literature predating \citet{1965ApJ...142..531F} that first voiced disagreement with Parker's derivation in Appendix~A.  
In retrospect, this criticism would have been invalid had Parker chosen to examine a solution with $\delta T$ being spatially constant.
Interestingly, \citet{1970ApJ...160..659D} recognized that \eqref{eq:deltaToft} corresponds to the original solution obtained by \citet{1953ApJ...117..431P}.  
However, he did not regard this solution as distinct from TI.  
The important point is that uniform temperature fluctuations are not stabilized by a conductive heat flux, so this instability persists even when TI modes are completely stabilized by thermal conduction.  Despite uniform temperature fluctuations being the $k=0$ limit of a sinusoidal perturbation, we explicitly verify in Appendix~B that \eqref{eq:deltaToft} does not correspond to the long wavelength limit of a TI mode.
\par

\subsection{CC modes versus TI modes}
Consider a plasma subjected to a random distribution of perturbation amplitudes and wavenumbers.\footnote{We hereafter confine our attention to catastrophic cooling only, although we discuss the possibility of catastrophic heating in \S{4}.}
As shown in \S{3}, typical parameterizations of the heating rate that can balance the cooling rate of gas in collisional ionization equilibrium result in a wide parameter space where the plasma is simultaneously isobarically and isochorically unstable. 
Two questions then come to mind:
(1) how do we distinguish between TI modes and the uniform temperature fluctuations leading to isochoric catastrophic cooling (referred to as simply CC modes in what follows)?;
(2) given that the isobaric growth rate is often faster than the isochoric one, is it still possible for CC modes to dominate the evolution of the system?
\par

If we confine our attention to a 1D local simulation of a homogeneous plasma that was initialized with a random set of perturbations, it is clear how to answer question (1).   
Let the domain extend from $-L$ to $L$ and have periodic boundary conditions. Then there is a Fourier series representation of the initial temperature profile given by
\begin{equation}
T(x) = \avg{T(x)} + \sum_{m=1}^\infty \left( a_m \cos k_m x + b_m \sin k_m x \right),
\label{eq:Fseries}
\end{equation}
where $k_m = m\pi/L$ are individual wavenumbers, $a_m$ and $b_m$ are the Fourier coefficients (in this case, the perturbation amplitudes corresponding to $k_m$), and
\begin{equation}
\avg{T(x)}  = \f{1}{2L} \int_{-L}^L T(x)\,dx
\label{eq:Tavg}
\end{equation}
is the average of the temperature profile.  
The CC mode has $\delta T < 0$ and can be simply computed from the constant component as $\delta T = \avg{T(x)} - T_{eq}$.  
The rest of the Fourier decomposition will consist of a superposition of MHD waves and condensation modes susceptible to TI.  
\par

We answer question (2) in the affirmative by assuming that the perturbations to saturate first are those that become nonlinear first.  We then define the timescale for reaching nonlinearity, $t_{non}$, through the equation
\begin{equation}
A \exp(\sigma t_{non}) = \f{1}{2},
\label{eq:tnon}
\end{equation}
where $\sigma$ is the instability growth rate.
That is, we assume that exponential growth is halted once the amplitude of a perturbed quantity grows to half the background value of that quantity.  
For TI modes, we set $A = |\delta \rho|/\rho_{eq}$ (for background density $\rho_{eq}$)
and take the growth rate to be that of the isobaric limit, namely 
$\sigma = -c_p^{-1}(\pdtext{\mathcal{L}}{T})_p$ \citep[see][]{1970ApJ...161...55D}, where $c_p$ is the specific heat at constant pressure.  For CC modes, $A = |\delta T|/T_{eq}$ and by \eqref{eq:deltaTeqn}, $\sigma = -c_V^{-1}(\pdtext{\mathcal{L}}{T})_\rho$.  By \eqref{eq:tnon} then, 
\begin{equation}
t_{non} = 
    \begin{dcases*}
        \f{c_p}{|(\pdtext{\mathcal{L}}{T})_p|} \ln\f{2|\delta \rho|}{\rho_{eq}} & \text{   (TI modes)}\\
        \f{c_V}{|(\pdtext{\mathcal{L}}{T})_\rho|} \ln\f{2|\delta T|}{T_{eq}}    & \text{   (CC modes).}
    \end{dcases*}
\end{equation}
We expect a CC mode to dominate any particular TI mode whenever $(t_{non})_{CC} < (t_{non})_{TI}$, which corresponds to a lower bound on the magnitude of $\delta T$.  Defining the ratio of specific heats $\gamma = c_p/c_V$ and the ratio of the isobaric and isochoric asymptotic growth rates,
\begin{equation}
R = \f{1}{\gamma} \f{(\pdtext{\mathcal{L}}{T})_p}{(\pdtext{\mathcal{L}}{T})_\rho},
\label{eq:R}
\end{equation}
which was shown by \citetalias{2019ApJ...875..158W} to be the basic parameter governing the various regimes of TI, this bound can be written
\begin{equation}
|\delta T| >  \f{1}{2}\left(\f{2|\delta\rho|}{\rho_{eq}}\right)^{1/R}T_{eq}.
\end{equation}
Returning to the above conceptual 1D simulation,
there will be one $(a_m,k_m)$ or $(b_m,k_m)$ combination that corresponds to the TI mode with the shortest $t_{non}$.  Denoting this mode by $\delta \rho_{max}$, we can express the above inequality as the following upper bound on the mean temperature: 
\begin{equation}
\avg{T(x)}  <  \left[1 - \f{1}{2}\left(\f{2|\delta\rho_{max}|}{\rho_{eq}}\right)^{1/R}\right] T_{eq} .
\label{ineq:Tavg}
\end{equation}
\par

To use this bound, note that \citetalias{2019ApJ...875..158W} showed that $R \geq 1$ when a plasma is thermally unstable by both the isochoric and isobaric criterion---unless the isentropic instability criterion is also satisfied, implying the overstability of acoustic modes, in which case $1/\gamma < R < 1$ (see their Appendix~B). 
Because $R$ is the ratio of the maximum isobaric and isochoric growth rates, the CC mode grows faster than all condensation modes in plasmas with $0<R<1$.
A typical case might be $R\approx 1$ and $|\delta\rho_{max}|/\rho_{eq} = 0.1$, putting the cutoff at $0.9\,T_{eq}$.  Below this temperature, we would expect the CC mode to control the thermal state of the gas, meaning it can overcome the runaway heating taking place in the under-dense portion of all TI modes, resulting in the catastrophic cooling of the entire plasma --- a `single-phase' rather than `two-phase' outcome. 
\par

\section{Isochoric instability}
As \citet{1995ASPC...80..328B} commented, the isochoric instability criterion as given by \ineqref{eq:isochoric} is much more difficult to satisfy compared to the isobaric one, $(\pdtext{\mathcal{L}}{T})_p < 0$, and this has proved true for cooling functions derived from photoionization equilibrium \citep[see e.g.][]{1997ApJ...478...94H,2009MNRAS.393...83C,Dannen19}.  However, it turns out to be easily satisfied for cooling functions based on collisional ionization equilibrium (CIE), as first noted by \citet{1971SoPh...19...86G} for the CIE-based radiative loss function of \citet{1969ApJ...157.1157C}. 
\par

\citetalias{2019ApJ...875..158W} derived a simple graphical criterion for judging if there are regions on the equilibrium curve with $(\pdtext{\mathcal{L}}{T})_\rho < 0$: the slope of those regions when plotted on the $\log T$ vs. $\log n$ plane must be positive.  This result follows immediately from the identity 
\begin{equation}
    \left(\pd{T}{\rho}\right)_\mathcal{L} = - \f{(\pdtext{\mathcal{L}}{\rho})_T}{(\pdtext{\mathcal{L}}{T})_\rho}.
\end{equation}
Multiplying both sides by $\rho/T$ (and introducing a mean particle mass $\bar{m} = \rho/n$ to exchange $\rho$ for the number density $n$) gives 
\begin{equation}
    \left(\pd{\log T}{\log n}\right)_\mathcal{L} = - \f{\rho}{T}\f{(\pdtext{\mathcal{L}}{\rho})_T}{(\pdtext{\mathcal{L}}{T})_\rho}.
\end{equation}
The logarithmic derivative is just the slope of the equilibrium curve (the contour where $\mathcal{L}$ vanishes), and its sign depends on the isochoric temperature derivative in the denominator on the right hand side.
Thus, this slope is positive whenever $(\pdtext{\mathcal{L}}{T})_\rho < 0$ provided $(\pdtext{\mathcal{L}}{\rho})_T > 0$, which is the statement that the plasma should cool when the density is increased at a fixed temperature --- a property that is difficult to \textit{not} satisfy.
\par

\citetalias{2019ApJ...875..158W} also showed that if $(\pdtext{\mathcal{L}}{\rho})_T > 0$ and $(\pdtext{\mathcal{L}}{T})_\rho < 0$, it must be the case that the isobaric instability criterion, $(\pdtext{\mathcal{L}}{T})_p < 0$, is also satisfied.  
We used this property in \S{2.2} to constrain the value of $R$ appearing in condition \eqref{ineq:Tavg}.  
The reason we refer to `TI modes' rather than just condensation modes is due to yet another property of TI: when the plasma is both isochorically and isobarically unstable, all three varieties of condensation modes (what \citetalias{2019ApJ...875..158W} called the entropy mode and the fast and slow isochoric modes) can exist for $R < 1/3$.  For $R > 1/3$, just one condensation mode remains (the fast isochoric one) and its stability, like the CC mode, is set by the sign of the derivative $(\pdtext{\mathcal{L}}{T})_\rho$.  
The isobaric derivative $(\pdtext{\mathcal{L}}{T})_p$ is instead tied to an overstable acoustic wave.
\par

\subsection{CIE-based cooling functions}
For CIE-based cooling functions, the property $(\pdtext{\mathcal{L}}{\rho})_T > 0$ should probably be regarded as a physical constraint on the density-dependence of the coronal heating rate.
A widely used form of the net cooling source term from \eqref{eq:energy} in solar corona modeling is
\begin{equation}
\rho \mathcal{L} = n^2 \Lambda(T) - C p^a T^b,
\label{eq:rhoL}
\end{equation}
where $\Lambda(T)$ is the radiative loss function and $C$ is a constant.  
This specification of the heating rate dependence follows \citet{1988SoPh..117...51D}, who quoted values of $a$ and $b$ appropriate for the various heating mechanism dependencies identified by \citet{1978ApJ...220..643R}.  Solving \eqref{eq:rhoL} for $\mathcal{L}$ gives
\begin{equation}
\mathcal{L} = \f{\rho}{\bar{m}^2} \Lambda(T) - C_H \left(\f{\rho}{\rho_{eq}}\right)^{a-1} \left(\f{T}{T_{eq}}\right)^{a + b},
\label{eq:CIEfunc}
\end{equation}
where we have eliminated $n$ and $p$ in favor of $\rho$ and defined a new heating constant that is given by $C_H = (\rho_{eq}/\bar{m}^2) \Lambda(T_{eq})$, chosen according to the desired starting location $(\rho_{eq},T_{eq})$ along the equilibrium curve.  Taking the derivative with respect to $\rho$ at fixed $T$ gives 
\begin{equation}
    \left(\pd{\mathcal{L}}{\rho}\right)_T = \f{1}{\bar{m}^2} \Lambda(T) + (1-a) \f{C_H}{\rho_{eq}} \left(\f{\rho}{\rho_{eq}}\right)^{a-2} \left(\f{T}{T_{eq}}\right)^{a + b},
\end{equation}
showing that this derivative will always be positive provided $a \leq 1$.
The only heating mechanism discussed by \citet{1978ApJ...220..643R} that has $a > 1$ is Alfven wave damping (with $a = 7/6$), but that dependence may still keep $(\pdtext{\mathcal{L}}{\rho})_T > 0$.  In any case, the one-to-one correspondence between positive slopes and isochoric instability certainly holds for $a \leq 1$.
\par

\begin{figure}
  \centering
  \includegraphics[width=.75\textwidth]{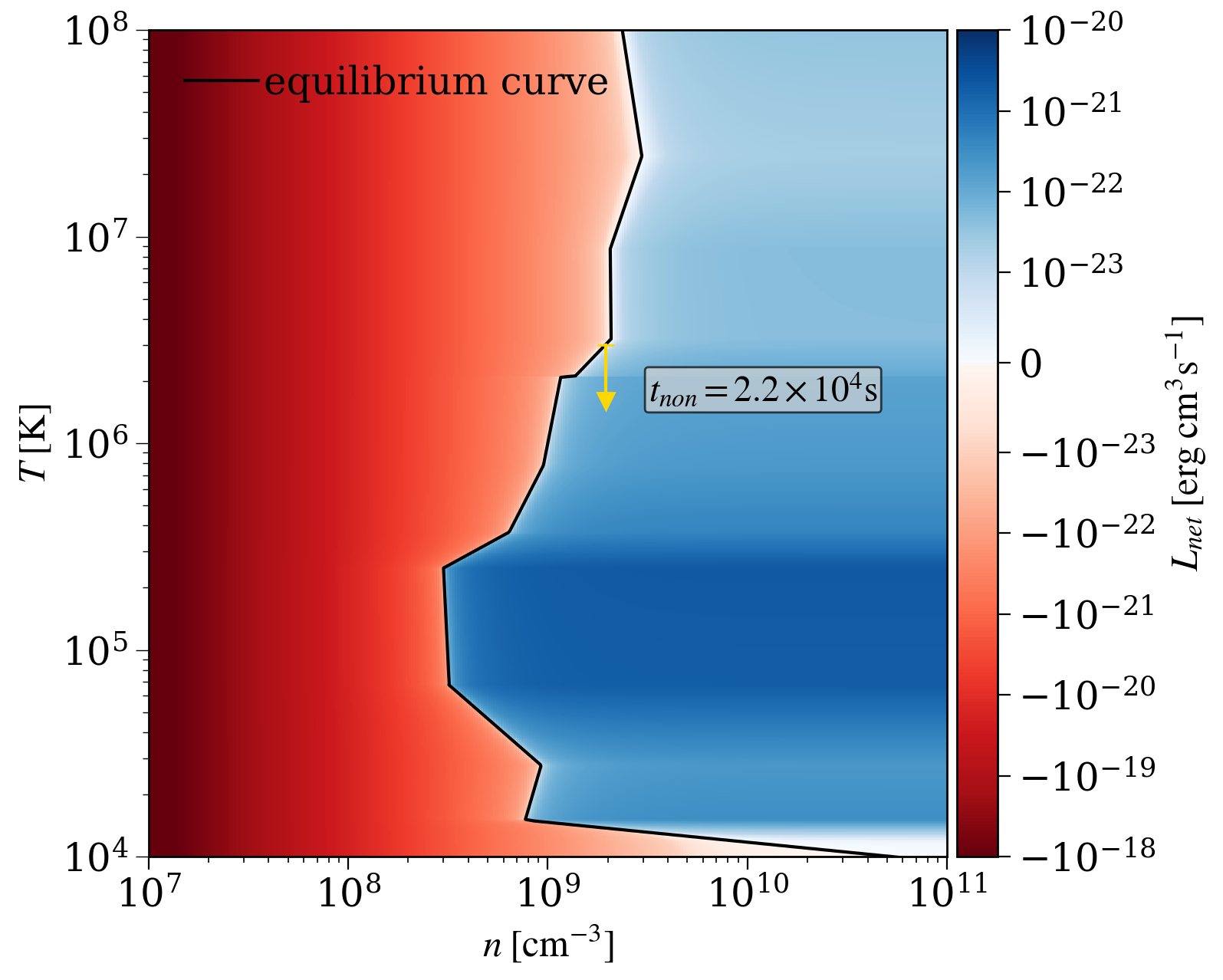}
  \caption{Temperature-density phase diagram for a plasma in CIE subject to a constant volumetric heating rate.  The colormap shows $L_{\rm{net}} = (\bar{m}/n)\mathcal{L}$ calculated using a piecewise powerlaw fit to the SPEX radiative loss function of \citet{2009A&A...508..751S}.  In blue (red) regions, $\mathcal{L} > 0$ ($\mathcal{L} < 0$), indicating net cooling (heating). 
  Overplotted is the contour $\mathcal{L}=0$; all positive-sloping segments of this equilibrium curve are both isobarically and isochorically unstable and plasma there will undergo catastrophic cooling instability in addition to TI.  The arrow points along an isochoric path and depicts the characteristic timescale for the catastrophic cooling mode to become nonlinear starting from an amplitude $\delta T = -10^3{\rm K}$ ($|\delta T|/T_{eq} = 4\times10^{-4}$).  The additional time to reach the stable equilibrium location at $T \approx 1.5\times 10^4{\rm K}$ is $\Delta t \approx 0.1\,t_{non}$.
  }
  \label{fig1}
\end{figure}

In Fig.~\ref{fig1}, we plot the equilibrium curve for the most common choice of heating parameters in numerical studies: $a = b = 0$, corresponding to a constant volumetric heating rate (see \eqref{eq:rhoL}).  
The portion of this curve with $3\times 10^5\rm{K} < T < 3\times 10^6\rm{K}$ has a positive slope and thus satisfies the isochoric instability criterion.  
The colormap makes this intuitively obvious: on negative sloping portions of the curve, a vertical displacement below equilibrium puts the gas into a (red) net heating region, where it will return to equilibrium.  
But on positive sloping portions, such as the one marked by an arrow, gas is unstable because it enters a (blue) net cooling region.
\par

That the overall occurrence of isochoric instability is not sensitive to the choice of $\Lambda(T)$ or the heating parameters is easily demonstrated.  
In Fig.~\ref{fig1}, we used a standard radiative loss function, one based on a piecewise powerlaw fit of the tabular function given in Table~2 of \citet{2009A&A...508..751S},\footnote{Specifically, we used the piecewise powerlaw calculated by \citet{Hermans_EPS_2021} and obtainable from \url{https://erc-prominent.github.io/team/jorishermans/}} which was calculated using the SPEX package \citep{2000adnx.conf..161K}.
\citet{2021A&A...655A..36H} compared this $\Lambda(T)$ with several other popular functions based on CIE (including \citet{2001ApJ...553..440K,2008ApJ...689..585C}); all curves showed a non-monotonic rise of the cooling rate to a peak around $10^5\,\rm{K}$, followed by non-monotonic fall as temperature increases.  
In other words, there are several temperature ranges for which $d\Lambda(T)/dT < 0$.  From \eqref{eq:CIEfunc} with $a + b = 0$, there is a very simple correspondence
between the slope of $\Lambda(T)$ itself and the value of the isochoric derivative, namely
\begin{equation}
    \left(\pd{\mathcal{L}}{T}\right)_\rho = \f{\rho}{\bar{m}^2} \f{d\Lambda(T)}{dT}
    \text{\hspace{25pt}($a + b = 0$)}.
\end{equation}
Thus, all temperatures for which $\Lambda(T)$ decreases with $T$ are isochorically unstable; this property of radiative loss functions has long been recognized \citep[see][and references therein]{1993ApJ...418L..25G}.

\par
Intuitively, when $d\Lambda(T)/dT < 0$ and the heating rate is constant, gas at fixed density that started to cool slightly will keep cooling at a higher rate.  
When $a + b \neq 0$, this simple correspondence between isochoric instability and the shape of the radiative loss function no longer holds, but that with positive slopes of the equilibrium curve on the $(\log T, \log n)$-plane still does.  
\par

For arbitrary $a$ and $b$,
\begin{equation}
    \left(\pd{\mathcal{L}}{T}\right)_\rho = \f{\rho}{\bar{m}^2} \f{d\Lambda(T)}{dT} -  (a + b)\f{C_H}{T_{eq}} \left(\f{\rho}{\rho_{eq}}\right)^{a-1} \left(\f{T}{T_{eq}}\right)^{a + b - 1},
\end{equation}
revealing that the heating term has a stabilizing contribution to the isochoric derivative when $a + b < 0$.  
Referring again to \citet{1978ApJ...220..643R}, only one of the identified dissipation mechanisms satisfies this: the conductive damping of Alfven modes ($a = 1/2$, $b = -1$).  
Hence, the occurrence of isochoric instability appears to be the norm for CIE-based cooling functions.  
\par

The most commonly used CIE-based cooling functions are known to satisfy the isochoric instability criterion \citep[e.g.] []{2011ApJ...737...27X,2015AdSpR..56.2738M, 2024A&A...688A.145J},
but we know of only one example of a phase diagram showing the equilibrium curve: that given in the Appendix of \citet{2021A&A...655A..36H}.  By itself, the equilibrium curve does not reveal the full extent of instability; to gauge stability away from equilibrium, a colormap of the net cooling function is also helpful, as in our Fig.~\ref{fig1}.\footnote{The quantity $L_{\rm{net}} \equiv (\bar{m}/n)\mathcal{L}$ is particularly useful because it has the same units as $\Lambda(T)$.}
Returning to this figure, it is clear that starting from temperatures above or below the equilibrium curve, a static, homogeneous plasma will not isochorically approach $T_{eq}$ unless $T_{eq}$ lies on a stable location.  
The question how plasma can occupy unstable locations in the first place is a very good one and deserves comment.  In local simulations, the only way is to set initial conditions with $T = T_{eq}$.  
In global simulations, the velocity field can allow flows to settle into unstable parameter space slightly off the equilibrium curve.  

For example, after \citet{Dannen20} uncovered TI operating in 1D Parker-like wind solutions (in calculations based on photoionization equilibrium), \citet{2021ApJ...914...62W} obtained steady state versions of these solutions upon reducing the amplitude of perturbations associated with the interpolation of the tabular source term.  
These studies demonstrated that low amplitude condensation modes have no chance of growing if gradients in the velocity field can stretch the modes before TI can amplify them.  
General flow fields therefore do not necessarily `avoid' linearly unstable parameter space unless the cooling timescale is much shorter than the dynamical timescale.
\par

\section{Summary and conclusions}
We have presented the following results: 
\begin{itemize}
\item Catastrophic cooling instability is a separate linear instability than TI that obeys the simple (and wavenumber-independent) dispersion relation $\sigma = -(1/c_V) (\pdtext{\mathcal{L}}{T})_\rho$.  Its associated eigenvector, denoted the CC mode, is simply the difference between the mean plasma temperature and the initial equilibrium temperature.
\item Because the CC mode cannot be stabilized by thermal conduction, it can lead to condensation formation in environments where the Field length exceeds characteristic perturbation wavelengths. 
\item The CC mode can dominate all thermally unstable condensation modes in the system when the mean plasma temperature obeys the upper bound given in \ineqref{ineq:Tavg}, in which case the outcome of a local simulation should be a single cold phase rather than a `two-phase' medium.
\item Cooling functions derived from collisionial ionization equilibrium are prone to violating the isochoric instability criterion that governs CC modes.  
\end{itemize}
\par

The second result references the Field length $\lambda_F$, as first defined by \citep{Begelman83}.
As for the third result, it is unclear what the outcome will be when the CC mode is dominated by only a few fast growing condensation modes.  
Provided the ratio $R$ defined in \eqref{eq:R} is greater than 1, such modes would have the shortest wavelengths allowed by linear theory, i.e. $\lambda \gtrsim \lambda_F$, so there may not be enough hot plasma produced in the underdense portions of these modes to overcome the bulk cooling by the CC mode. 
One would then expect a two-phase medium that is mostly cold gas with embedded pockets of hot gas (the remnants of the short wavelength condensation modes).  
If we then extrapolate this expectation when considering a local region within a global simulation, we arrive at a mechanism for producing larger condensations than are possible with isobaric TI modes.  
\par

Alternatively, in global simulations, the CC mode may only have enough time to partially cool a small region of background gas (much smaller than the density scale height $\lambda_\rho = \rho/|\nabla \rho|$ of a stratified atmosphere, for example) before the dynamics associated with either the surrounding plasma or the saturation of TI disrupt its growth.  This latter outcome may account for what \citet{2013ApJ...773...94M} labeled `complete' versus `incomplete' condensations in a parameter study of 1D coronal loop models \citep[see also][]{2018ApJ...855...52F}.
\par

Either of the above outcomes is expected to lead to oscillatory dynamics akin to the nonisobaric regime of TI \citepalias[see][]{2019ApJ...875..158W}. 
The distinguishing property of CC modes is the accompanying loss of pressure support, so as the perturbation becomes nonlinear, strong flows can develop as mechanical equilibrium becomes lost during the evolution away from thermal equilibrium.  
This nonisobaric dynamics may offer a natural explanation for the observed long intensity pulsations accompanying some types of coronal rain \citep{2022FrASS...920116A}.  

The overall tendency for the evolution to runaway from equilibrium leads us to suggest that the growth of CC modes is the mechanism underlying condensation formation in studies of the TNE model (see \S{1}).
In coronal loops, longitudinal TI modes should be completely stabilized by thermal conduction since most loops have values of $\lambda_F$ exceeding the loop length.  Thus, either TNE is a fully independent and nonlinear process for forming condensations, or there is a nonlocal counterpart to the CC mode acting in the simulations discussed in \S{1}.  A compelling description of what sounds more like a counterpart to a CC mode than to a TI mode was given by \citet{1985ApJ...291..798B}, who in describing their global eigenfunctions stated, `the form of the perturbation we have studied is such that $\delta T$ is maximum at the apex, and decreases monotonically toward the base $s=0$ (where the perturbation vanishes); a typical temperature eigenfunction\ldots has no nodes; only this type of perturbation can be unstable (i.e., perturbations with nodes are all stable).'
\par

Our analysis has centered on catastrophic cooling because our interests are in understanding coronal rain.
We could have instead focused on an instance of catastrophic heating by reversing the yellow arrow in Fig.~\ref{fig1} and considering a $\delta T > 0$ perturbation (i.e. a CH mode) of partially heated coronal plasma located on the equilibrium curve at $T \approx 3\times10^5\,\rm{K}$.
Since there is no stable equilibrium in this direction, the plasma temperature would rise to values in excess of $10^6\,\rm{K}$ where conductive losses could supply a balance.  
It would of course be remarkable if this local instability --- based on a single, simple ordinary differential equation --- is relevant to the solution of the famous coronal heating problem \citep[e.g.][]{2006SoPh..234...41K}.  
The important property to consider is its monolithic character: when this instability operates in a region of constant density, it will uniformly heat or cool that entire region.\footnote{Whether heating or cooling predominates is likely governed by the background gradients; TNE loops are more prone to coronal rain formation because the thermal conduction flux acts to cool the corona, so perturbations are naturally sent below the local equilibrium temperature set by $\mathcal{L} = 0$.}
By contrast, condensation growth through TI involves a local force, i.e. it is the pressure gradient between the overdense and underdense gas that drives perturbation growth, hence TI can only operate on the scales of individual modes.  
If catastrophic heating instability was to play any role at all, it would probably be in tipping the balance in favor of overall heating by continually amplifying the CH modes supplied by magnetic dissipation processes.

\appendix
\section{The early literature on TI}
Several authors concluded early on that Parker's demonstration of an instability arising from the heat equation alone \citep[see][]{1953ApJ...117..431P} could not actually manifest when solving the coupled equations of non-adiabatic, compressible MHD.
For example, referring to Parker's study as well as \citet{1956ApJ...123..299A}, \citet{1960ApJ...132..452W} states `one ought perhaps to consider the stability criterion of Athay \& Thomas and of Parker, as applying to the case of incompressible fluids, since arbitrary disturbances will affect the density as well as the temperature of the medium...'.  Upon viewing an arbitrary disturbance as consisting of a constant displacement above or below $T_{eq}$ in addition to a superposition of acoustic or condensation modes (see \S{2.2}), it becomes clear that TI governs only the latter modes, while the constant displacement $\delta T = \avg{T(x)} - T_{eq}$ is the mode of a separate instability altogether.  The instability acts to exponentially amplify the magnitude of $\delta T$ --- the departure of the mean temperature away from equilibrium --- and thus catastrophic cooling/heating instability is an appropriate name.  

Equations (1), (2), and (3) of \citet{1970ApJ...160..659D}'s important study showing that a thermally unstable stratified atmosphere is also convectively unstable are precisely \eqref{eq:Teqn}, \eqref{eq:deltaTeqn}, and \eqref{eq:isochoric} here.  He discounts this solution by stating the following:
\begin{quote}
  Weymann (1960) pointed out that the condition (3) for thermal instability applies only to isochoric perturbations. But density perturbations are to be expected, owing to the strong tendency of a fluid to remain in pressure equilibrium. A work term must therefore be added to equation (1), and the density dependence of the heat-loss function must be taken into account. 
\end{quote}
This is a true statement and leads one to the dispersion relation for TI, but it is not a valid reason to ignore the instability described by his starting equations.

\section{The long wavelength limit of TI}
Our claim has been that the CC mode does not correspond to the asymptotic limit of any particular TI mode (i.e. of a condensation mode or an overstable acoustic wave).  Intuitively, TI modes are `centered about equilibrium', whereas the CC mode is a finite displacement above or below $T_{eq}$.  Formally, we can eliminate consideration of the $k=0$ asymptotic limit of overstable modes by noting that their growth rates in this limit are always proportional to either $k$ or $k^2$ \citepalias[see Figure~2 from][]{2019ApJ...875..158W}.  Only the fast isochoric condensation mode has the same growth rate as the CC mode in this limit. We therefore just need to examine the analytic solution for condensation modes, which can be written as \citep[see][]{2023FrASS..1098135W}

\begin{equation}
\begin{split}
\rho(\gv{x},t) &= \rho_0 + A \rho_0 e^{\sigma t} \cos(\gv{k}\cdot{\gv{x}})  \\
v(\gv{x},t) &= v_0 - A v_c e^{\sigma t}\sin(\gv{k}\cdot{\gv{x}})  \\
p(\gv{x},t) &= p_0-A \rho_0 v_c^2 e^{\sigma t} \cos(\gv{k}\cdot{\gv{x}}),
\end{split}
\label{eq:linear_profiles}
\end{equation}
where $\rho_0$, $v_0$, and $p_0$ are the density, velocity, and pressure of the uniform background flow, $A$ is the perturbation amplitude, and $v_c\equiv \sigma/k$ (for wavenumber $k = |\gv{k}|$) is the solution to the following cubic dispersion relation:
\begin{equation}
v_c = - \f{k^{-1}}{c_V}\left(\pd{\mathcal{L}}{T}\right)_\rho \f{R + (v_c/c_{s,0})^2}{1 + (v_c/c_{s,0})^2}.
\label{eq:vcDR}
\end{equation}
Here, $c_{s,0} = \sqrt{\gamma p_0/\rho_0}$ is the sound speed in the background flow.  

The long wavelength limit of TI has been associated with isochoric instability in the literature because upon substituting back in for $\sigma = k v_c$ and taking the limit $k\rightarrow 0$, one obtains the isochoric growth rate $\sigma = -c_V^{-1}(\pdtext{\mathcal{L}}{T})_\rho$.  However, as explained by \citetalias{2019ApJ...875..158W} and \citet{2023FrASS..1098135W}, the long wavelength isochoric limit should instead be thought of as the highly nonisobaric regime of TI, characterized by condensation velocities $v_c$ exceeding the speed of sound.  Indeed, taking the limit $v_c \gg c_{s,0}$ in \eqref{eq:vcDR} more directly gives back $\sigma = -c_V^{-1}(\pdtext{\mathcal{L}}{T})_\rho$.  The isobaric regime of TI, on the other hand, has $v_c \ll c_{s,0}$, and \eqref{eq:vcDR} gives back $\sigma = -c_p^{-1}(\pdtext{\mathcal{L}}{T})_p$ instead.  In both limits, therefore, the condensation velocity scales as $v_c \propto k^{-1}$.  Referring to the analytic solution above, the only way to recover the CC mode is to demand $A \propto k^2$ so that both $A$ and $A v_c$ vanish as $k\rightarrow 0$, while $A v_c^2$ can remain finite.  Because $A$ is just an arbitrary constant amplitude in the theory of local TI, this analysis shows that the CC mode cannot be the $k=0$ limit of the isochoric condensation mode.  However, in a WKBJ analysis of TI in a stratified atmosphere, $A$ gets promoted to a position-dependent variable.  This suggests that in a nonlocal perturbation analysis, global counterparts to the CC mode can be found by taking the $k\rightarrow 0$ limit of dispersion relations for condensation modes; the temperature eigenmode plotted in Figure~2 of \citet{1985ApJ...291..798B} might be one such example. 

\begin{acks}

We thank Fan Guo for discussions about magnetic heating mechanisms and James Klimchuk for thought provoking email correspondence. 
TW acknowledges the DOE Office of Science's \href{https://science.osti.gov/wdts/suli}{SULI program}, 
through which Tess Boland was employed at LANL and wrote numerous analysis routines using \href{https://www.sympy.org/en/index.html}{SymPy} for a related project, which made working with piecewise cooling functions much simpler.  
TW additionally thanks Patrick Kilian for conversations about kinetic processes in the solar atmosphere, while AS would like to thank Craig Johnston, James Klimchuk, and Dmitrii Kolotkov for stimulating discussions during the 11\textsuperscript{th}  Coronal Loops Workshop, as well as Sean Sellers for his insights.

\end{acks}

 \begin{authorcontribution}
T.W. wrote primary manuscript, A.S. reviewed derivations and prepared figures.  All authors reviewed the manuscript.
 \end{authorcontribution}

\begin{fundinginformation}
This work was supported by the U.S. Department of Energy through the Los Alamos National Laboratory. 
Los Alamos National Laboratory is operated by Triad National Security, LLC, for the National Nuclear Security Administration of U.S. Department of Energy (Contract No. 89233218CNA000001).
\end{fundinginformation}


\end{document}